\documentstyle[11pt,paspconf,psfig]{article}

\newcommand{\gtap}{\mathrel{\hbox{\rlap{\lower.55ex \hbox {$\sim$}}
                   \kern-.3em \raise.4ex \hbox{$>$}}}}
\newcommand{\ltap}{\mathrel{\hbox{\rlap{\lower.55ex \hbox {$\sim$}}
                   \kern-.3em \raise.4ex \hbox{$<$}}}}

\begin{document}

\title{X-rays from old star clusters}

\author{Frank Verbunt}
\affil{Astronomical Institute, Utrecht University, Postbox 80.000,
   3508 TA Utrecht, The Netherlands; email verbunt@phys.uu.nl}

\begin{abstract}
A brief overview is given of X-ray observations of old clusters. 
Most X-ray sources in old open clusters are interacting binaries, 
formed via evolution of a primordial binary, and emitting
X-rays because of magnetic activity; however, a sizable fraction of the
cluster sources is not well understood, including some of the most luminous 
ones. Globular clusters 
appear to contain fewer magnetically active X-ray sources than expected 
if one scales from old open clusters by mass.

\end{abstract}


\section{Introduction}

The comparison of stellar clusters with different ages has contributed much 
to our understanding of the evolution of single stars.
Large numbers of binaries are being discovered and studied in stellar clusters 
via radial velocity studies (as reviewed by Mermilliod 1997), and
we may hope to learn about the evolution of {\em binaries} as well.
Another reason to study binaries in clusters is the close interaction
between the evolution of the binaries and the evolution of the cluster
as a whole (see review by Hut et al.\ 1992).
For example, close encounters between single stars and binaries change the
velocity distribution of the cluster stars. Exchange encounters, in which
a single star encounters a binary and releases one binary star by taking its
place, may lead to strange binaries which would not -- or not as frequently --
be formed via the evolution of an isolated binary.

In this paper I review the role in these studies
of X-ray observations of old ($\gtap 1\,$Gyr) clusters. In Section\,2
I discuss what types of objects emit X-rays and how these objects -- very
often binaries -- are formed, and give an overview
of all X-ray observations of old open clusters (for an earlier review,
see Belloni 1997).
In Section\,3 the bright and dim X-ray sources in globular clusters
are discussed, and a comparison with the sources in old open clusters
is made in Section\,4.

\section{X-rays from old open clusters}

\subsection{A brief overview of close binary evolution}

To prepare for our discussion of the X-ray sources in old open
clusters, we make a brief detour to binary evolution, illustrated in Figure\,1.
Consider a primordial binary of two main-sequence stars in a relatively
close orbit (Fig.\,1-Ia).
Unless pre-main-sequence circularization has occurred, the binary will
in general have an eccentric orbit.
Loss of angular momentum drives the two stars together, and at some point
tidal interaction starts: the larger (i.e.\ more massive) star first
is spun-up to be brought into co-rotation with the orbit at periastron.
Subsequently the orbit is circularized. Rapid rotation engenders chromospheric
and coronal activity, and with it, X-ray emission (Ib).
When loss of angular momentum is severe, or the inital binary very close,
the two stars may come into contact (Ic) and may even merge into
a single 'FK\,Com' star (Id). Both contact binaries and FK\,Com
stars are magnetically active X-ray emitters.

\begin{figure}[tbp]
\centerline{\psfig{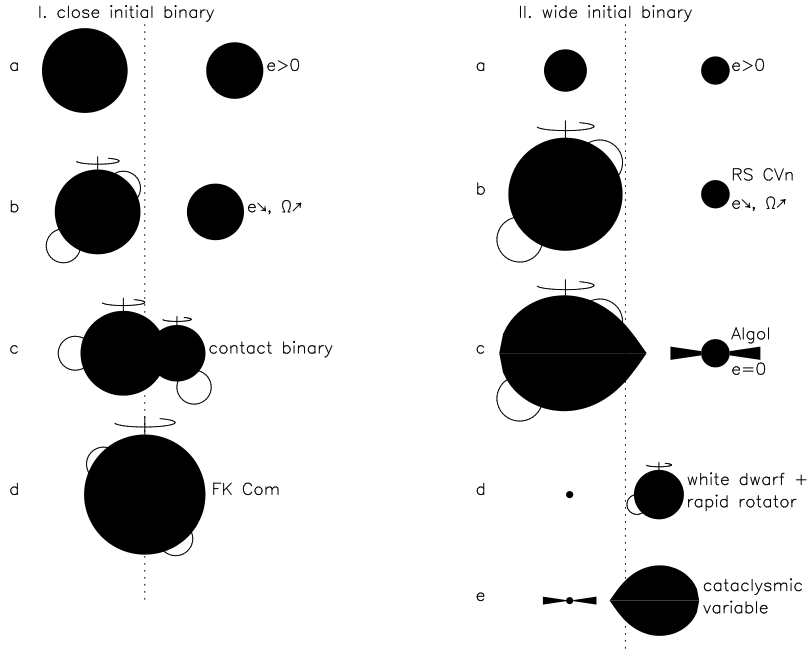}}
\caption{Schematic outline of the evolution of binaries
with an initially close orbit (left) and with an initially wide
orbit (right). For details see text.
\vspace*{0.2cm}}
\end{figure}

When the initial binary is wide (Fig.\,1-IIa), loss of angular momentum has
no effect, and tidal forces only come into play once one of the binary
stars -- the more massive one -- evolves into a (sub)giant. The subgiant
is then brought into corotation, and the orbit circularized; the binary
becomes an RS CVn system, in which the rapidly rotating star emits
X-rays (IIb). When the giant fills its Roche lobe, mass transfer starts and
the binary is an Algol system (IIc), until the complete envelope of the
giant has been transferred, leaving a white dwarf into orbit around a
companion. The white dwarf is undermassive, because mass transfer has
interrupted the ordinary giant evolution; and its companion has been
brought into rapid rotation because of the angular momentum it accreted
with the mass (IId). The companion to the white dwarf may evolve to
fill its Roche lobe either because it expands into a giant, or because
loss of angular momentum lets the orbit shrink. The ensuing mass transfer
then dominates the luminosity, and the binary has become a cataclysmic 
variable. (Note, however, that the typical progenitor of a cataclysmic variable
has a rather wider orbit than characteristic Algols.)

With the exception of the initial binaries, all the evolutionary stages
shown in Fig.~2 are X-ray sources. Most of them emit X-rays due to
magnetic activity induced by rapid rotation; only in the case of the
cataclysmic variable the X-rays are due to the mass accretion onto the
white dwarf.

\subsection{Optical identification of X-ray sources}

\begin{figure}[tbp]
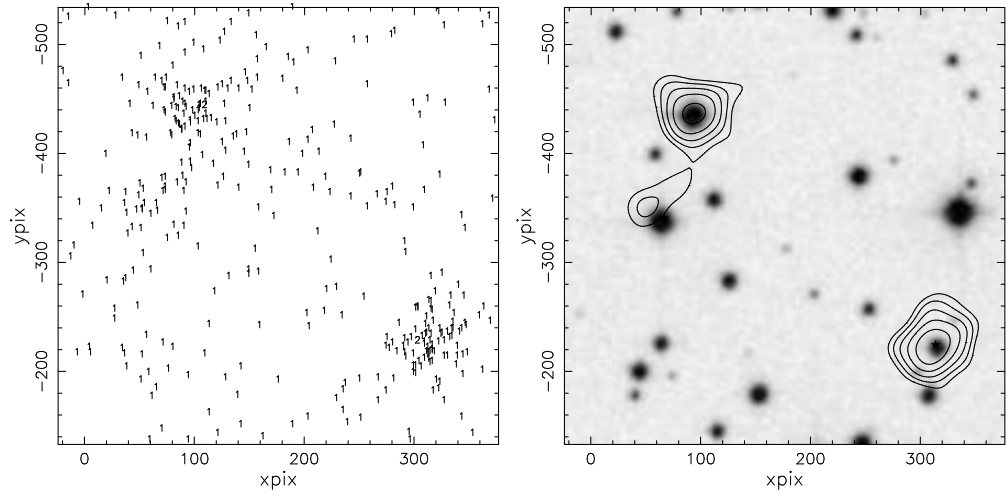

\centerline{
     \begin{minipage}[b]{2.6in}
          \psfig{figure=verbuntf2a.ps,width=2.6in,clip=t}
     \end{minipage}
     \begin{minipage}[b]{2.6in}
          \psfig{figure=verbuntf2b.ps,width=2.6in,clip=t}
     \end{minipage}
}
          \caption[]{Individual X-ray photons detected on a part of
the ROSAT PSPC image of M\,67 (left), turn into sources when the image is 
smoothed with a $\sigma$$=$$6''$$=$$12$\,pixels Gaussian (right, contours).
Overlaying the smoothed image on an optical image we identify the
two bright sources (S\,1082 above and S\,1063 below); the third
source (S\,1072) shows an offset, which may be explained as due to
the small number of photons of that source. At this offset, however,
chance coincidences of X-ray positions with optical objects cannot
be ignored.
}
\end{figure}

To illustrate the optical identification of X-ray sources in old open
clusters, we discuss some details of the X-ray analysis of M\,67. In Fig.\,2a
we show the distribution of counts on the ROSAT detector near S\,1082
and S\,1072. It is seen that most pixels are empty, some contain a
single count, and some two.  If we smooth the picture and then plot X-ray
contours, three sources show up (Fig.\,2b).  Overlaying these sources
on an optical image, we find that the two brighter ones coincide
closely with S\,1082 and S\,1063. These identifications may be
considered secure.  We also find that the X-ray source that Belloni et
al.\ (1998) identify with S\,1072 has a maximum which is offset with
respect to the optical position.  The number of photons that defines
this X-ray source is small, and the X-ray center has a 90\%\
uncertainty radius of 9$''$; the offset between the X-ray center and
S\,1072 is at this radius, and the identification is feasible.  We
note from the figure that a circle with a 9$''$ radius located
arbitrarily in the frame has a finite probability of hitting an
optical object. Belloni et al.\ (1998) estimate that 1 of the 12
identifications of X-ray sources with member binaries of M\,67 may be
due to chance coincidence; most identifications, possibly all,
are correct.  In IC\,4651 two X-ray sources have multiple optical
counterparts of comparable brightness in their error circle; this 
indicates that the probability of chance coincidence in this
cluster is high.

\begin{table}
\caption[o]{Overview of the X-ray sources in old open clusters.
For each cluster I list the age, the
X-ray detection limit (in erg/s in
the 0.1-2.4\,keV energy range), the number $N$ of X-ray sources
(tentatively) identified with optical cluster members, and of these
the numbers of spectroscopic binaries with eccentricities zero, larger than
zero, and unknown, respectively; of contact binaries, FK\,Com
systems, blue stragglers; and of other X-ray sources whose nature is ('oth')
or is not ('?') understood. References are a) Belloni et al.\ 1998,
b) Belloni \&\ Verbunt 1996, c) Belloni \&\ Tagliaferri 1998,
d) Belloni \&\ Tagliaferri 1997.
}
\begin{flushleft}
\begin{tabular}{lcccccccccccl}
cluster & age & limit & $N$ & \multicolumn{3}{c}{SB} & CB & FK & BS &
oth & ? & ref\\
        & (Gyr) & $10^{30}$ &     & $e$$=$$0$ & $e$$>$$0$ & ? \\
\\
NGC\,188 & 6   & 4.0 & 2  &   &   & 1 &   & 1 &   &   &   & a \\
M\,67    & 4   & 0.8 & 25 & 3 & 3 & 4 & 2 &   & 1 & 3 & 3 & a \\   
NGC\,752 & 2   & 0.7 & 8  & 4 &   &   &   &   & 1 & 1 &   & b \\  
IC\,4651 & 1.5 & 1.5 & 6  &   & 1 &   &   &   & 1 &   &   & c \\
NGC\,6940 & 1  & 1.0 & 4  & 2 & 1 &   &   &   &   &   &   & d \\
\end{tabular}
\end{flushleft}
\end{table}

\subsection{X-ray sources in old open clusters}

Table~1 summarizes the X-ray observations of old open clusters, and the
identifications of X-ray sources with optical objects in different categories.
The richest harvest has been obtained for M\,67. The optical counterparts of
the X-ray sources in M\,67 are located throughout the colour-magnitude diagram,
both near the isochrone for single stars, and away from it (Figure\,3).
Most sources are known binaries. The smaller number of
X-ray sources in NGC\,188 is most likely due to the rather higher
detection limit in that cluster. The smaller numbers of X-ray sources
in NGC\,752, IC\,4651 and NGC\,6940 is due in part because of the smaller
number of stars in these clusters; and in part perhaps to the absence of
a well-developed (sub)giant branch in these younger clusters (Figure\,3).
Old clusters were ignored in early X-ray studies of open clusters because
old stars do not emit much X-rays -- or so it was thought. Ironically,
the least luminous sources detected in old clusters are brighter
than the most luminous sources in young open clusters (see Randich, these 
proceedings).
The lower limit in the old clusters is set by the detection limit; less 
luminous sources are certainly present.

\begin{figure}[tbp]
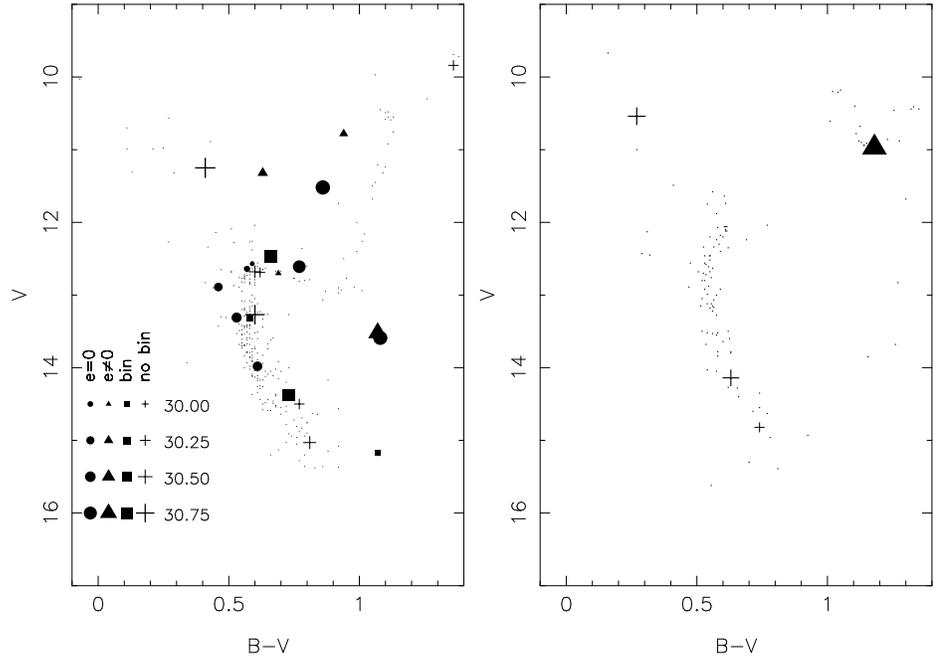

\centerline{
     \begin{minipage}[b]{2.4in}
          \psfig{figure=verbuntf3a.ps,width=2.4in,clip=t}
     \end{minipage}
     \begin{minipage}[b]{2.4in}
          \psfig{figure=verbuntf3b.ps,width=2.4in,clip=t}
     \end{minipage}
}
\caption{Colour magnitude diagram of M\,67 (left, after Belloni et al.\ 1998) 
and IC\,4651 (right). Special
symbols indicate stars detected in X-rays:\ circles, triangles and squares
indicate binaries with circular, eccentric and unknown orbits, respectively;
$+$ stars for which no evidence for binarity was found.
The symbol size indicates the (logarithm of the) X-ray luminosity
(between 0.1-2.4\,keV, in ergs/s). 
}
\end{figure}

We discuss the X-ray sources in old open clusters on the basis of Table\,1, 
starting with those for which the X-ray emission is readily understood. 
Circular binaries (SB $e$$=$$0$ 
in Table\,1) are X-ray sources because of tidal interaction.
Those in NGC\,752 have orbital periods of 0.41, 1.01, 1.45 and 1.95\,d,
those in M\,67 of 2.66, 7.65 and 10.06\,d; these binaries are probably
of the type shown in Fig.\,1-Ib. The circular binaries in NGC\,6940
have orbits of 54.2 and 82.5\,d, and are probably of the type
shown in Fig.\,1-IIb.
Two contact binaries (CB in Table\,1; Fig.\,1-Ic) are found in M\,67, and an
FK\,Com star (FK in Table\,1) in NGC\,188.
Other X-ray sources which we understand ('oth' in Table\,1)
are a cataclysmic variable
(Fig.\,1-IIe), a hot single white dwarf, and a triple system in which a close
binary of the type shown in Fig.\,1-Ib is kept at finite eccentricity
by a third companion in a wide orbit in M\,67; and a rapidly rotating
star -- probably rotating rapidly because of tidal interaction
in a close binary -- in NGC\,752. 

Turning to the sources for which the X-ray emission is less readily 
understood, we note that a blue straggler (BS in Table\,1) has been detected
in three of the five old clusters. In each case, the blue straggler is
among the brightest X-ray sources in the cluster. Multicolour photometry of 
the ones in M\,67 and IC\,4651 indicates that they are binaries
(Landsman et al.\ 1998, Anthony-Twarog \&\ Twarog 1987), and radial
velocities indicate that the one in NGC\,752 is a binary 
(Latham, cited in Belloni 1997), but no
period is known. It is not clear why the blue stragglers emit X-rays.
The brightest X-ray source in M\,67 is similar to the binary depicted in
Fig.\,1-IId (Landsman et al.\ 1997, see also Verbunt \&\ Phinney 1995), 
except that the companion to the white dwarf is a slow rotator (van den
Berg et al.\ 1999, \&\ these proceedings). 
It is seen at $V$$=$$11.52$, $B$$-$$V$$=$$0.88$
in Fig.\,3a. If rapid rotation is taken to be pre-requisite for X-ray
emission, we do not understand this source. The binary AY\,Cet is
rather similar to this system, and also is a relatively bright X-ray
source. The X-ray emission of this system is probably related to its
earlier history, which included a phase of mass transfer; magnetic
activity is present even in the absence of rapid rotation, but we
do not know why.
We note that X-ray emission of giants in general is not
well related to rotational velocity (Van den Berg et al.\ 1999).

The next two brightest X-ray sources in M\,67 are both binaries, 
located below the subgiant branch, several magnitudes above the
main sequence (at $V$$\simeq$$13.5$, $B$$-$$V$$\simeq$$1.1$
in Fig.\,3a); no single star or binary can be located there
according to our current understanding. Both binaries show H\,$\alpha$
emission and emission in the cores of the Ca\,H\&K lines, and thus
appear to be magnetically active X-ray sources. S\,1063 has a 18.39\,d
orbit with eccentricity $e=0.22$; S\,1113 has a circular 2.82\,d orbit.
Both are photometric variables. We do not currently understand their
evolutionary status (Van den Berg et al.\ 1999, \&\ these proceedings).

Several spectroscopic binaries with eccentric orbits have also been 
detected in X-rays (SB $e$$>$$0$ in Table\,1). The binary in IC\,4651
has an orbit of 75\,d, and a relatively small eccentricity ($e$$=$$0.09$,
Mermilliod et al.\ 1995); perhaps it is in the process of tidal 
circularization. One binary in M\,67 has an orbit of 31.78\,d and a fairly
high eccentricity of $e$$=$$0.664$ (Mathieu et al.\ 1990); a photometric period
has been detected at 4.88\,d (Gilliland et al.\ 1991), which is exactly the 
corotation period at periastron, strongly suggestive of tidal interaction.
The three other eccentric binaries detected in X-rays have very long
orbital periods, 697.8 and 1495\,d in M\,67 and 3595\,d (!) in NGC\,6940
(Mermilliod \&\ Mayor 1989).
Such wide orbits exclude tidal interaction, as confirmed by the eccentric
orbits. One wonders whether these stars could be triple systems,
i.e.\ the giant star that we observe has a close binary as a companion.
Triple systems are not uncommon: one X-ray source in M\,67 is known to be 
a triple, and  one of the X-ray sources in the same ROSAT field of view as 
M\,67 is identified with HD\,75638, a bright F star not related to the 
cluster (Belloni et al.\ 1998), which is a triple star (Nordstr\"om et
al.\ 1997).

Finally, several spectroscopic binaries with as yet unknown orbits have
been detected as X-ray sources; considering that half of the X-ray
detected binaries with known orbits are unusual, we may expect more surprises
when these binaries are studied further.

\section{X-ray sources in globular clusters}

Twelve bright ($L_x\gtap 10^{36}$\,erg/s) X-ray sources have been found in
globular clusters; they are neutron stars accreting matter
from a companion star. Four of these are bright only occasionally,
i.e.\ they are soft X-ray transients.
The number of bright sources in the whole galaxy is
of order 100. Since globular clusters contain only $\sim$$0.001$ of the
mass of our Galaxy, the high incidence of bright X-ray sources in them points
to a formation process for such sources which operates preferably in 
globular clusters. This process has been identified with the occurrence
of close encounters between stars: if a neutron star passes close to
another star it may be captured tidally; alternatively a neutron star
may eject a binary star and take its place in an exchange encounter
(for a review, see Hut et al.\ 1992).
The orbital period is known for five bright X-ray sources in globular 
clusters; for two of them this period is so small (11 and 13 -- or 20 --
minutes, respectively; Homer et al. 1996) that the mass donor must be a 
white dwarf. 
No such systems are known in the galactic disk, again pointing to a 
special formation mechanism in globular clusters.

\begin{figure}[tbp]
\centerline{
     \begin{minipage}[b]{2.7in}
          \psfig{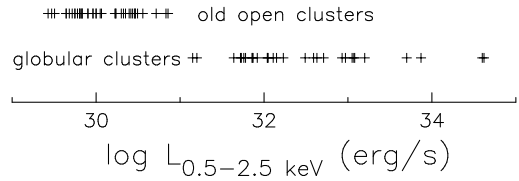}
     \end{minipage}
     \begin{minipage}[b]{2.7in}
          \caption{\hspace*{-1em}
X-ray luminosity distributions of X-ray sources in old open clusters and dim 
sources in globular clusters (luminosities from references listed in Table\,1 
and from Johnston \&\ Verbunt 1996).
}
     \end{minipage}
}
\end{figure}

The nature of the less luminous X-ray sources discovered in globular
clusters, with $L_x\ltap10^{35}$\, erg/s, is less clear.
About thirty dim sources are now known distributed over twenty clusters,
their X-ray luminosity distribution is shown in Figure\,4.
Many more must be present, below current detection limits. 
The luminosities of the dim sources are such that transients in their 
low-luminosity state, cataclysmic variables, RS\,CVn systems, and recycled 
radio pulsars all are possible counterparts.
The only secure identification is with a recycled radio pulsar in M\,28:
the radio period of 3.054\,ms is detected also in X-rays (Danner et al.\ 
1997).
Various other sources have been tentatively identified with cataclysmic
variables.
Figure\,5 illustrates the problems that optical identification of X-ray 
sources in the cores of globular clusters entail.The central region of
47\,Tuc contains five X-ray sources, and several thousand stars: given
the limited accuracy of the X-ray positions, a possible counterpart is
always found.
Even if one limits oneself to blue and/or variable stars only,
the probability of chance coincidence is still appreciable.
Thus, X-ray sources C and D in Fig.\,4 can tentatively be identified
with cataclysmic variables at nearby positions, but chance coincidence
cannot be excluded (Verbunt \&\ Hasinger 1998).

NGC\,6397 is another cluster containing multiple dim X-ray sources in its
core, for which cataclysmic variables have been suggested as counterparts
(Cool et al.\ 1995). The sources are so close together, however, that
the whole core of the cluster is covered by their error circles, i.e.\ any
cataclysmic variable (or other star) in the cluster core of necessity
falls within an X-ray error circle.
Clearly, secure identification can be obtained only if either the 
positional accuracy is improved (as may be done with AXAF), or
if the same periodicity is discovered in the X-ray source and in the
suggested optical counterpart.

\begin{figure}[tbp]
\centerline{
     \begin{minipage}[b]{2.7in}
          \psfig{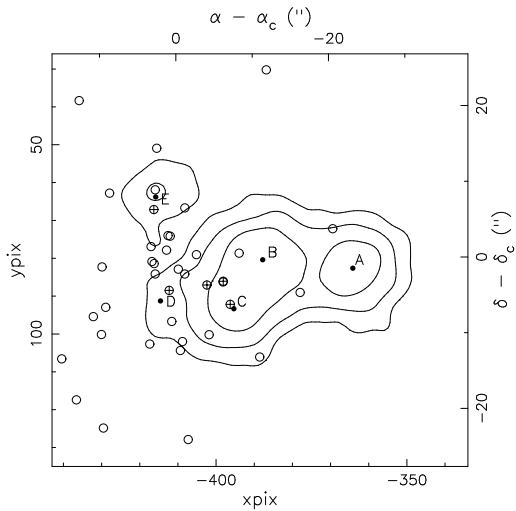}
     \end{minipage}
     \begin{minipage}[b]{2.7in}
          \caption{\hspace*{-1em}
X-ray contour levels of the core of 47~Tuc, with the
best locations of five X-ray sources A-E ($\bullet$). Open circles indicate 
blue stragglers and variable stars listed by Geffert et al.\ (1997); these
are on an optical coordinate system, which may be shifted with respect to
the contours by up to 2$''$. Among the suggested counterparts
(indicated with $\oplus$) two cataclysmic variables are good candidates
for X-ray sources C and D. After Verbunt \&\ Hasinger (1998).
}
     \end{minipage}
}
\end{figure}

A statistical analysis of the currently known dim sources in globular
clusters, taking into account the detection limit for each cluster,
shows that the number of sources $N$ at luminosity $L_x$ in a cluster
scales with the density $\rho_c$ and the mass $M_c$ of the cluster core 
as (Johnston \&\ Verbunt 1996):
$$ d N(L_x)\propto M_c{\rho_c}^{0.5}\times {L_x}^{-1.5} dL_x $$
This implies that the total luminosity of each cluster core is dominated
by a small number of bright sources rather than by a large number
of unresolved very faint sources; in agreement with observation.
If the only process of importance is the rate of close encounters between
stars in the cluster core, one expects a proportionality of the
source numbers $N\propto{\rho_c}^2\propto M_c\rho_c$; for primordial
binaries the expected proportionality is with $M_c$. The fact that the
observed proportionality lies between these two indicates that the
population of dim sources is a mix of primordial binaries evolved into
X-ray sources (e.g.\ RS\,CVn systems and cataclysmic variables)
and X-ray sources formed via close encounters (X-ray transients and
recycled radio pulsars; some cataclysmic variables).

\section{Comparison open and globular clusters; prospects}

In comparing the old open clusters with the globular clusters, we note
that the sources in globular clusters detected so far are brighter than
the brightest sources found in old open clusters (see Fig.\,4).
Most of the currently known X-ray sources in globular clusters are the
result of past stellar encounters in the cluster cores.
Most of the currently know X-ray sources in old open clusters may have
evolved from primordial binaries.
Some X-ray sources in old open clusters -- including all four brightest
sources in M\,67, and the brightest source in IC\,4651 -- are not well 
understood. In general however it may be stated that X-ray observations
are very efficient in picking out binaries in old open clusters which are
currently interacting.  

The brightest X-ray sources in old open clusters are an FK Com system in
NGC\,188 and a slightly eccentric binary in IC\,4651, both with
$L_x({\rm 0.1-2.4\,keV})>10^{31}$\,erg/s. The four brightest sources
in M\,67 each have $L_x({\rm 0.1-2.4\,keV})\simeq0.75\times10^{31}$\,erg/s.
There is no obvious reason why such systems couldn't exist in globular
clusters as well.
However, if we assume that a globular cluster typically has a hundred to
a thousand times more mass than an old open cluster and that the total
X-ray luminosity scales with mass, we predict a that each globular cluster
has an X-ray luminosity well in excess of $10^{33}$ to $10^{34}$\,erg/s,
in contrast to observation (Verbunt 1996).
We investigate this more closely in Figure\,6, where we compare the
X-ray luminosity to mass ratio of globular clusters with that of
M\,67.

\begin{figure}[tbp]
\centerline{
     \begin{minipage}[b]{2.2in}
          \psfig{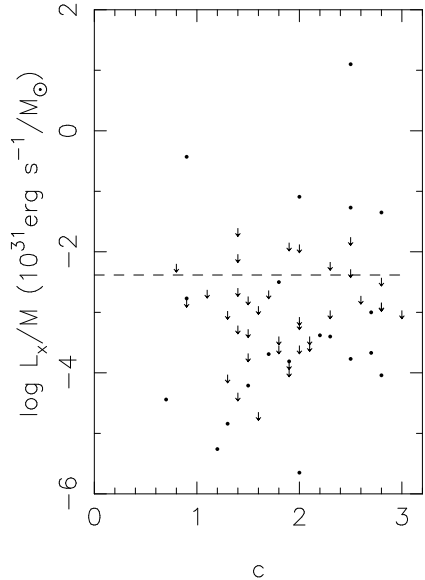}
     \end{minipage}
     \begin{minipage}[b]{2.9in}
          \caption{\hspace*{-1em}X-ray luminosity to mass ratio
of globular clusters ($\bullet$ detection, $\downarrow$ upper
limit) as a function of the concentration of the cluster 
$c\equiv \log r_c/r_t$, where $r_c$ and $r_t$ are the core and tidal
radius. 
X-ray luminosities in the 0.5-2.5\,keV range, from Verbunt et al.\ (1995), 
Johnston \&\ Verbunt (1996); cluster mass and concentration from
Pryor \&\ Meylan (1993).
The dashed line shows the $L_x/M$ ratio for M\,67 (assumed 
mass 724$\,M_{\odot}$, Montgomery et al.\ 1993).
}
     \end{minipage}
}
\end{figure}

It is seen that most globular clusters have a lower X-ray light to mass
ratio than M\,67, some by several orders of magnitude.
This may indicate  that binaries are destroyed efficiently in the
cores of globular clusters -- as suggested by the X-ray luminosity
function discussed in Sect.\,3. Since some of the brightest X-ray
sources have very short orbital periods and the brightest source
in NGC\,188 is a single star, this explanation may not be sufficient.
An alternative possibility is that M\,67 has lost
a larger fraction of its mass -- in particular single, non X-ray emitting
stars -- than the typical globular cluster.
It is certainly true that the fraction of binaries amongst the giants
in M\,67 is larger than in the well-studied globular clusters.

Ottmann et al.\ (1997) find that binaries of population II are less luminous
X-ray emitters than their population I counterparts; such a finding would
also explain the dearth of X-ray detections of magnetically active sources 
in globular clusters. It should be noted that the sample on which Ottmann
et al.\ base their conclusion is rather small; and that it includes just one
binary with an evolved primary (viz.\ HD\,6286; Pasquini \&\ Lindgren 1994).

The search for X-ray sources in globular clusters has concentrated
on their cores, because of confusion problems with background
sources away from the core. It may be worthwhile to estimate the
X-ray luminosities of globular cluster out to their tidal radius,
and then re-do Fig.\,6.

The prospects for further work are excellent.
AXAF and XMM are ready for launch. AXAF will provide more accurate
positions and more secure identifications, XMM will probe clusters to much
lower X-ray luminosities, so that magnetically active binaries become
detectable in globular clusters.
Theoretically the first descriptions of complete cluster evolution,
in which stellar and dynamical evolution is combined, are becoming
possible with special-purpose computers (Hurley et al, these proceedings).

\acknowledgments
I am grateful to Marten van Kerkwijk for enabling me to produce Fig.\,2;
to Luca Pasquini for directing me to the Ottmann et al.\ (1997) paper, and 
to Maureen van den Berg for comments on the manuscript.
I made extensive use of the SIMBAD database.


\begin{references}

\reference
Anthony-Twarog, B. \& Twarog, B. 1987, \aj, 94, 1222

\reference
Belloni, T. 1997, Mem. Soc. Astron. Ital., 68, 993

\reference
Belloni, T., Tagliaferri, G. 1997, \aap, 326, 608

\reference
Belloni, T., Tagliaferri, G. 1998, \aap, 335, 517

\reference
Belloni, T., Verbunt, F. 1996, \aap, 305, 806

\reference
Belloni, T., Verbunt, F., Mathieu, R. 1998, \aap, 339, 431

\reference
Cool, A., Grindlay, J., Cohn, H., Lugger, P., Slavin, S. 1995, \apj, 439, 695

\reference
Danner, R., Kulkarni, S., Saito, Y., Kawai, N. 1997, Nature, 388, 751

\reference
Geffert, M., Auri{\`e}re, M., Koch-Miramond, L. 1997, \aap, 327, 137

\reference
Gilliland, R., Brown, T., Duncan, D., Suntzeff, N., Lockwood, G., Thompson, D.,
  Schild, R., Jeffrey, W., Penprase, B. 1991, \aj, 101, 541

\reference
Homer, L., Charles, P., Naylor, T., van Paradijs, J., Auri\`ere, M.,
  Koch-Miramond, L. 1996, \mnras, 282, L37

\reference
Hut, P., McMillan, S., Goodman, J., Mateo, M., Phinney, S., Pryor, C., Richer,
  H., Verbunt, F., Weinberg, M. 1992, \pasp, 104, 981

\reference
Johnston, H., Verbunt, F. 1996, \aap, 312, 80

\reference
Landsman, W., Aparicio, J., Bergeron, P., Di~Stefano, R., Stecher, T. 1997, 
  \apj, 481, L93

\reference
Landsman, W., Bohlin, R., Neff, S., O'Connell, R., Roberts, M.S.and~Smith, A.,
  Stecher, T. 1998, \aj, 116, 789

\reference
Mathieu, R., Latham, D., Griffin, R. 1990, \aj, 100, 1859

\reference
Mermilliod, J.-C. 1997, Mem. Soc. Astron. Ital., 68, 853

\reference
Mermilliod, J.-C., Andersen, J., Nordstr{\"o}m, B., Mayor, M. 1995, \aap, 299,
  53

\reference
Mermilliod, J.-C., Mayor, M. 1989, \aap, 219, 125

\reference
Montgomery, K., Marschall, L., Janes, K. 1993, \aj, 106, 181

\reference
Nordstr{\"o}m, B., Stefanik, R., Latham, D., Andersen, J. 1997, \aaps, 126, 21

\reference
Ottmann, R., Fleming, T., Pasquini, L. 1997, \aap 322, 785

\reference
Pasquini, L., Lindgren, H.  1994, \aap 283, 179

\reference
Pryor, C., Meylan, G. 1993,
\newblock in S. Djorgovski, G. Meylan (eds.), Structure and Dynamics of
  Globular Clusters, {\em ASP Conference Series\/} Vol.~50, p.~357

\reference
van~den Berg, M., Verbunt, F., Mathieu, R. 1999, \aap, 347, 866

\reference
Verbunt, F. 1996,
\newblock in P. Hut, J. Makino (eds.), Dynamical evolution of star clusters,
  IAU Symp. 174, Kluwer Academic Publishers, Dordrecht, p.~183

\reference
Verbunt, F., Bunk, W., Hasinger, G., Johnston, H. 1995, \aap, 300, 732

\reference
Verbunt, F., Hasinger, G. 1998, \aap, 336, 895

\reference
Verbunt, F., Phinney, E. 1995, \aap, 296, 709

\end{references}
\end{document}